# Global Assessment of Oil and Gas Methane Ultra-Emitters

**Authors:** T. Lauvaux[1], C. Giron[2], M. Mazzolini[2], A. d'Aspremont[2,3], R. Duren[4,5], D. Cusworth[6], D. Shindell[7], P. Ciais[1]

**Affiliations:**

[1] Laboratoire des Sciences du Climat et de l'Environnement, IPSL, Univ. de Saclay, Saclay, France.

[2] Kayrros, Paris, France.

[3] CNRS & DI, Ecole Normale Supérieure, Paris, France.

[4] University of Arizona, Office of Research, Innovation and Impact, Tucson, AZ, USA

[5] Carbon Mapper, San Francisco, CA, USA

[6] Jet Propulsion Laboratory, California Institute of Technology, Pasadena, CA, USA

[7] Nicholas School of the Environment, Duke University, Durham, NC, USA.

*Correspondence to: thomas.lauvaux@lsce.ipsl.fr

**Abstract**:
Methane emissions from oil and gas (O&G) production and transmission represent a significant contribution to climate change. These emissions comprise sporadic releases of large amounts of methane during maintenance operations or equipment failures not accounted for in current inventory estimates. We collected and analyzed hundreds of very large releases from atmospheric methane images sampled by the TROPOspheric Monitoring Instrument (TROPOMI) over 2019 and 2020 to quantify emissions from O&G ultra-emitters. Ultra-emitters are primarily detected over the largest O&G basins of the world, following a power-law relationship with noticeable variations across countries but similar regression slopes. With a total contribution equivalent to 8-12% of the global O&G production methane emissions, mitigation of ultra-emitters is largely achievable at low costs and would lead to robust net benefits in billions of US dollars for the six major producing countries when incorporating recent estimates of societal costs of methane.

**One Sentence Summary:** Ultra-emitters from oil and gas production amount 8-12% of the global oil and gas methane emissions, offering actionable and cost-effective means to mitigate the contribution of methane to climate change.

**Intro:**

As the second-most important contributor to global warming, methane ($CH_4$) has continued to accumulate in the atmosphere by 50 Tg.yr$^{-1}$ over the last two decades, primarily due to increases in agricultural activities, waste management, coal, and Oil and Gas (O&G) production (Saunois et al., 2020; Jackson et al., 2020). Large discrepancies between atmospheric inversions, bottom-up inventories and biogeochemical models remain largely unexplained (Kirschke et al., 2013; Nisbet et al., 2014; Saunois et al., 2016, 2020). This complicates attribution of the recent global rise in atmospheric methane to an anthropogenic or biogenic source or a possible decline in the atmospheric OH radical sink (Rigby et al., 2017; Zhao et al., 2020) and/or to changes in biogenic and anthropogenic sources (Nisbet et al., 2019). Evidence of a large under-estimation of the fossil sources was suggested by the recent analysis of $^{14}CH_4$ isotopic ratios (Hmiel et al., 2020). Representing a quarter of anthropogenic emissions alone, emissions from O&G production activities have increased from 65 to 80 TgCH4.yr$^{-1}$ in the last 20 years (IEA, 2020). This rapid increase imperils the success of the Paris Agreement (Nisbet et al., 2020). Anthropogenic emissions trends are partly explained by the increase in shale gas production in the US, which is soon to be followed by large shale reserves currently under-exploited in China, Africa, and South America (IEA, 2013). While O&G emissions from national inventories have been widely underestimated by conventional reporting (Alvarez et al., 2018), airborne imagery surveys have confirmed the omnipresence of intermittent emissions, distributed according to a power law (Frankenberg et al., 2016; Duren et al., 2019; Cusworth et al., 2021) with a right-hand tail caused by very large O&G leaks, unintended or not, often referred to as *super-emitters* (Zavala-Araiza et al., 2015).

Until recently, observation-based $CH_4$ emission quantification efforts were restricted regionally to short duration (few weeks) aircraft surveys (Karion et al., 2015), or the deployment of in situ sensor networks (Lyon et al., 2020). Global efforts were limited by the sparse sampling of coarse-resolution $CH_4$ column retrievals, such as the GOSAT mission (Maasakkers et al., 2019). More routine and higher spatially-resolved emission quantification was made possible by the ESA Sentinel 5-P satellite mission carrying the TROPOspheric Monitoring Instrument (TROPOMI, launched 2018; Veefkind et al. 2012). TROPOMI samples daily $CH_4$ column mole fractions over the whole globe at moderate resolutions (5-7 km) revealing multiple individual cases of unintended very large leaks (e.g. Pandey et al., 2019) and regional basin-wide anomalies (Schneising et al., 2020; Barré et al., 2021). Here, we systematically examine this unique dataset over the globe, which represents the first opportunity to statistically characterize visible ultra-emitters of $CH_4$ from O&G activities across various basins. By nature, reducing these ultra-emitters using Leak Detection and Repair (LDAR) strategies provides an actionable and cost-efficient solution to emission abatement (Mayfield et al., 2017).

Detectable $CH_4$ enhancements from single point sources is limited by the TROPOMI instrument sensitivity (5-10ppb; Hu et al., 2018), by the overlap of multiple plumes from closely-located natural gas facilities (e.g. in the Permian basin), and by complex spatial gradients from remote sources affecting background conditions (cf. Supp. Info.). Rapidly varying meteorological conditions require sufficiently robust approaches, especially with curved $CH_4$ plume structures for which common mass balance methods are too simplistic (Varon et al., 2018). We addressed this problem by applying an automated plume detection algorithm and quantified the associated emissions using the Lagrangian particle model HYSPLIT (Stein et al., 2015) driven by meteorological reanalysis products for each detected plume enhancement (>25 ppb averaged over several pixels, cf. Supp. Info.) over the whole globe. The detection threshold is adjusted to only capture statistically significant enhancements within highly variable backgrounds (cf. Supp. Info.). Finally, we estimated the potential reductions along with abatement costs for various countries, to determine effective gains at national levels.

**Results:**

The number of detections of large $XCH_4$ enhancements around the world, each associated with an ultra-emitter, totals more than 1,800 single observed anomalies over two years (2019-2020), a large fraction of them located over Russia, Turkmenistan, the United States (excluding the Permian basin where regional enhancements comprise many small to medium emitters), the Middle East and Algeria (Fig. 1). Detections vary in magnitude and number between 50 to 150 per month, most of them corresponding to O&G production facilities (about two thirds of the detections, or ~1,200) while ultra-emitters from coal, agriculture and waste management only represent a relatively small fraction (33%) of the total detections (cf. SI). Ultra-emitters attributed to O&G infrastructures appear along major pipelines and over most of the largest O&G basins representing more than 50% of the total onshore natural gas production over the globe (IEA, 2021). Offshore emissions remain invisible to TROPOMI, hence excluded from our analysis (cf. Supp. Info.).

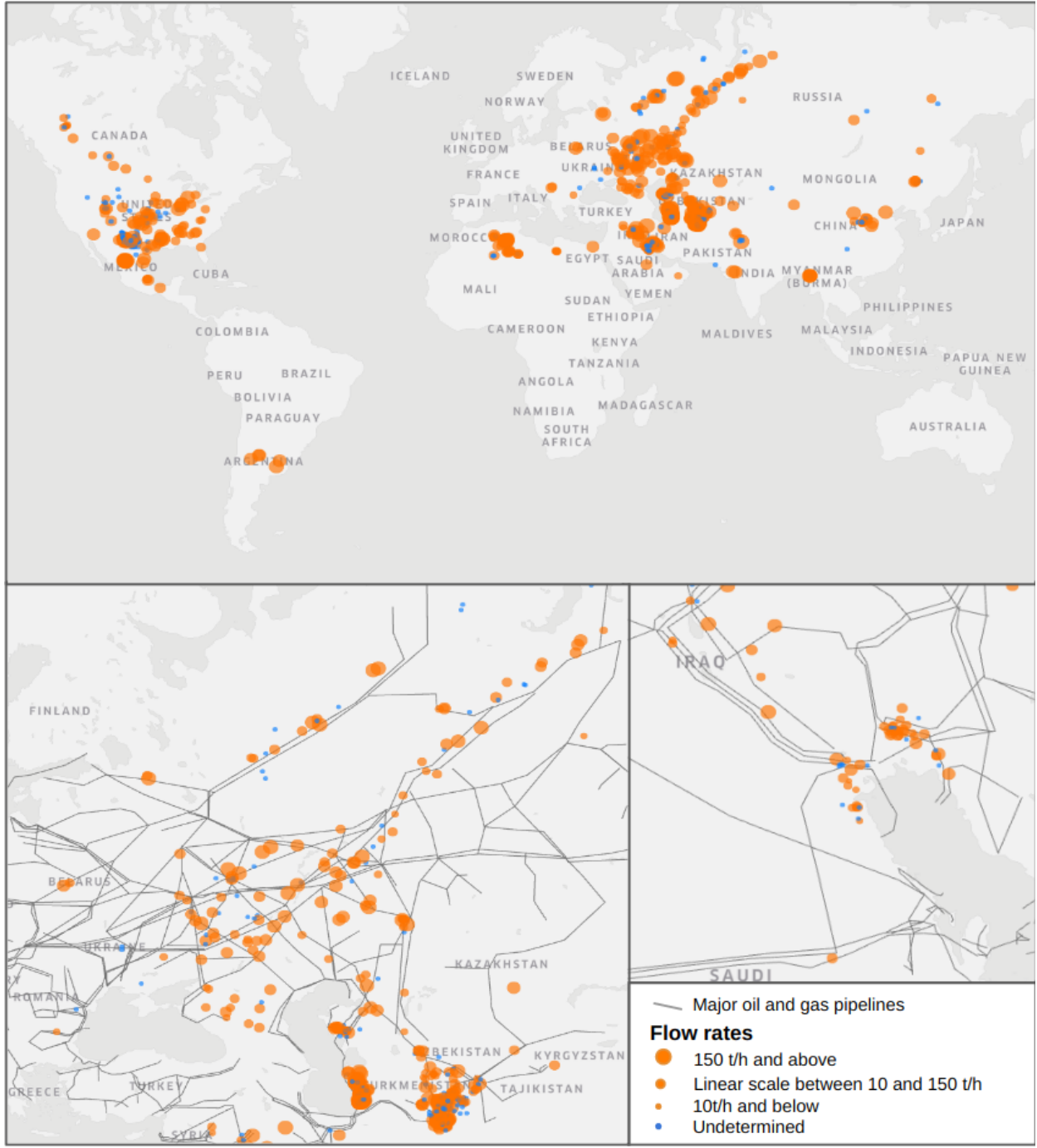

**Figure 1**: Global map of the ~1,200 O&G detections from TROPOMI over the years 2019 and 2020 (upper panel), zoomed-in over Russia and Central Asia (lower left panel) and over the Middle East (lower right panel) including the main gas pipeline (dark grey). Circles are scaled according to the magnitude of the ultra-emitters. Undetermined sources are indicated in blue. Map credit: MapBox.

Estimated emissions from O&G ultra-emitters rank highest for Russia with 1.5 MtCH$_4$.yr$^{-1}$, followed by Turkmenistan, the United States (excl. Permian basin), Iran, Kazakhstan and Algeria (Fig. 2a.). As leak duration varies while S5-P provides only snapshots, each leak duration was determined either based on observed duration as implied by the plume length (advection time) or setting a 24-hour duration when consecutive images confirmed the presence of the same anomaly over multiple days (Fig. 2a). Leaks lasting several days are adjusted by coverage loss, hence set to 24 hours (cf. Supp. Info.). Two additional scenarios were constructed to define the upper and lower bounds of durations using i) a systematic 24-hour duration, or ii) based on the length of the observed plumes (cf. Supp. Info.). The loss of coverage due to clouds albedo or aerosols was quantified by adjusting for the number of observed days compared to the full period length (cf. Supp. Info.). Uncertainties were quantified by a negative binomial probability function (Student, 1907; cf. Supp. Info.). We illustrate this adjustment in (Fig. 2a), large for some countries (e.g. Russia), by subsampling the coverage over Turkmenistan (originally 118) with the lowest coverage observed over a country (i.e. 22). After adjustment, estimated emissions fall within 2% of the original estimate and estimated uncertainty (1.26 $MtCH_4$) matches the full statistical test on the interval 0.96-1.6 $MtCH_4$ (Fig. 2 e.). Based on adjusted emissions, O&G ultra-emitter estimates represent 8-12% of O&G $CH_4$ emissions from national inventories (fig. 2b), a contribution not included in current inventories (Alvarez et al., 2018).

As one of the largest natural gas reserves of the world (~20 trillion cubic meters, ranking 4$^{th}$ in the world based on IEA), Turkmenistan is likely to see its O&G $CH_4$ emissions double simply because of ultra-emitters (Fig 2b.). Ultra-emitters are also relatively large in Russia, Iran, Kazakhstan and Iran representing between 10 to 20% of annual reported emissions. The United States revealed fewer ultra-emitters (5% of the annual inventory emissions) but we excluded the Permian basin (about 10% of the US natural gas production) due to the large basin-wide $XCH_4$ enhancement which obscures single detections (de Gouw et al., 2020). A recent study estimated at 2.7 $TgCH4.yr^{-1}$ the O&G emissions from the Permian using TROPOMI (Zhang et al., 2020), which represents 35% of the US O&G production emissions from the whole-US top-down estimate (Alvarez et al., 2018). Assuming infrastructure and maintenance operations are similar over the Permian and the rest of the US, the relatively small fraction of ultra-emitters should remain valid for the entire country. Middle Eastern countries like Iraq or Kuwait correspond to even fewer detections (31) possibly thanks to fewer accidental releases and/or more stringent maintenance operations. The detection limit of ultra-emitters is around 25 $tCH_4.h^{-1}$ while the largest events reach several hundred tons per hour with associated plumes spanning hundreds of kilometers. However, ultra-emitters from any oil and gas basin of the world follow unequivocally a power-law distribution (Fig. 2d.) which implies that if the power-law coefficients are well-defined, ultra-emitters scale directly with smaller emitters. To establish this relationship over a broader range of emissions, the power-law of smaller emitters (from 0.1 to $10tCH_4.h^{-1}$) observed in high-resolution airborne spectrometer images with AVIRIS-NG (Duren et al., 2019) was combined with the one of S5-P emitters revealing similar regression parameters (slope of 1.9-2.3; Fig. 2 c.). The actual number of ultra-emitters varies by country (Fig. 2 d.) but the relationship between the number of sources and their magnitudes remains similar in the range of

0.1 to 300 tCH$_4$.h$^{-1}$ over two gas basins of the US. Very small leaks (<100 kgCH$_4$.h$^{-1}$) mostly caused by nominal operations (i.e. pneumatic devices) might fall onto a different relationship (Omara et al., 2016), while larger leaks are mostly accidental or related to specific maintenance operations (Conley et al., 2016). Overall, the total fraction of CH$_4$ emissions from ultra-emitters remains difficult to quantify accurately due to the lack of observations of smaller emitters, but their relative contribution compared to known sources is non-negligible and thus offers a cost-efficient and actionable opportunity to reduce CH$_4$ emissions while natural gas production increases steadily by about 3% per year (IEA data).

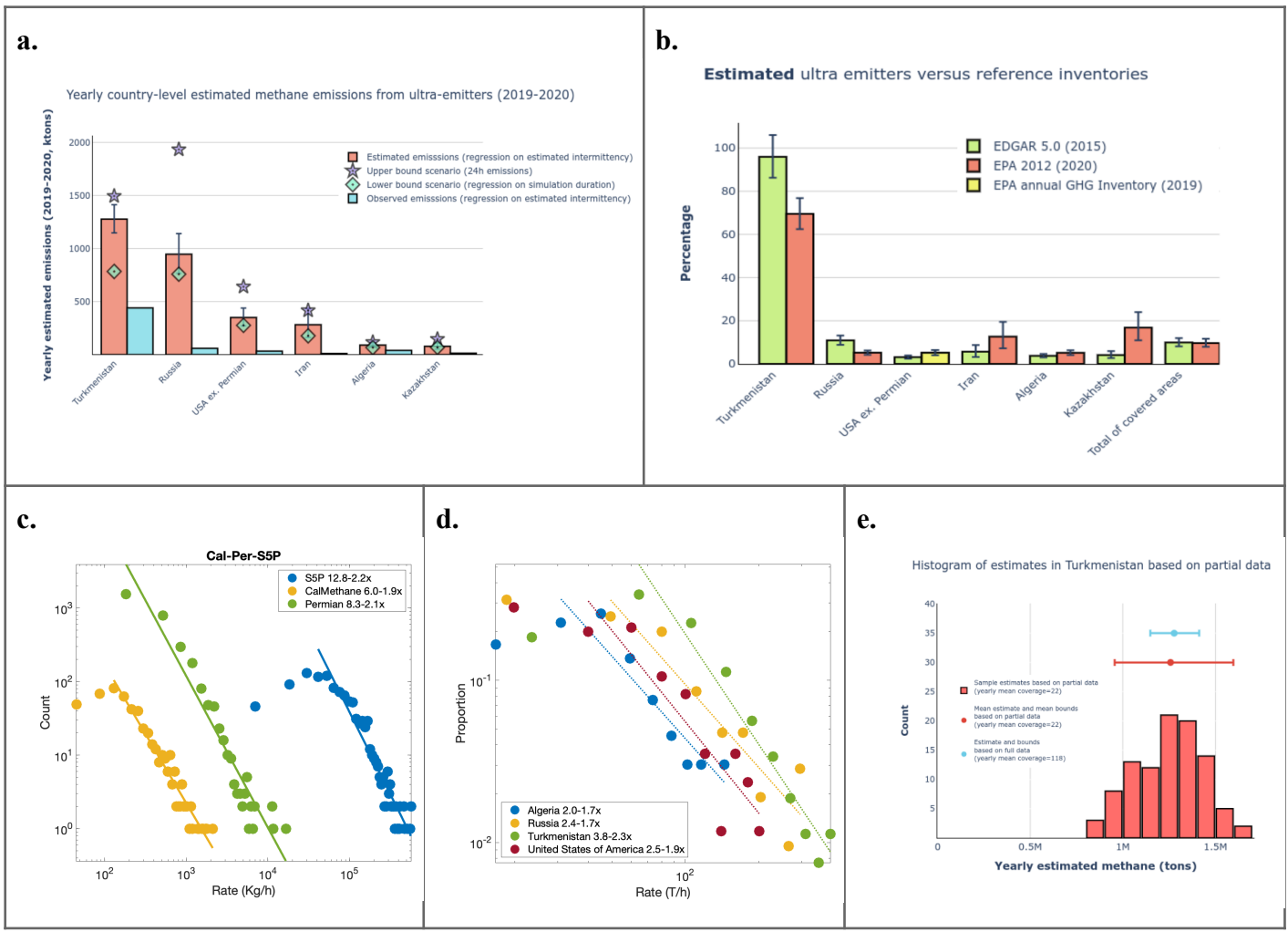

**Figure 2**: Country-level emissions from O&G ultra-emitters over the years 2019-2020 observed and estimated (adjusted for leak duration and coverage loss) together with two extreme leak duration scenarios (panel a.); Relative fraction of the estimated ultra-emitters to two national-scale methane inventories, EDGAR 5.0 and EPA (panel b.); Distribution of super-emitters from AVIRIS-NG and GAO campaigns over 2 years in California and two months in Texas (Duren et al., 2019; Cusworth et al., 2021)

and from 2-year Sentinel 5-P data (log-log scale; panel c.); same for S5-P only over four different countries (panel d.); and distribution of estimated emissions from sub-sampled S5-P detections compared to estimated emissions from full set for Turkmenistan (panel e.). EPA emissions (panel b) correspond to the latest 2012 global inventory extrapolated to 2020, except for the US (most recent EPA annual GHG inventory for 2019 (EPA, 2021)). Permian basin emissions were removed following the same methodology as in Zhang et al., 2020 (~1Mt/y).

We evaluate the industry spending required to eliminate those methane emissions based on analyses of mitigation costs recently produced by several groups: the International Energy Agency (IEA, 2021), the US Environmental Protection Agency (US EPA, 2019), and the International Institute for Applied Systems Analysis (IIASA; Höglund-Isaksson et al., 2020). All costs are evaluated in 2018 US$ per tonne methane. Briefly, we first analyze marginal abatement cost curves developed by these groups at the national level (regional level for IIASA) and excluding valuation of environmental impacts. As large emissions are expected to be related to upstream operations or long-distance transport of fuels, we exclude local distribution networks in the IIASA analysis which separates those sources. The IEA analysis provides separate cost estimates for high emission sources, whereas the other two do not. Those sources are expected to be more cost-effective to mitigate than average sources, however, and indeed the IEA estimates for our six countries of interest show costs ~$110-300 less than the average cost of mitigation in the O&G sector in those countries. We therefore evaluate average mitigation costs within the O&G sector for EPA and IIASA analyses screening for the subset of measures costing less than $600 per tonne. This same threshold was recently used to define 'low cost' controls (UNEP/CCAC, 2021), and would correspond to ~US$ 21 per tonne of carbon dioxide equivalent if converted using the IPCC Fifth Assessment Report's GWP100 value of 28 that excludes carbon-cycle feedbacks). Averaged across these mitigation analyses, spending is net positive in Iran (~$60 per tonne), whereas it is net negative in all other high-emitting countries with net savings of around $100-150 per tonne in Russia, Kazakhstan and Turkmenistan, about $250 per tonne in the US, and $400 per tonne in Algeria, though values vary greatly across the available analyses (Fig. 3a).

Examining the total spending required to eliminate the high emission sources in each country, there is a large spread across the available analyses. The analyses show the largest average expenditure in Iran, at $16 million, but a range of -$30 to 95 million across the analyses. Results for the US are more robust in that all show a net savings, but the values still vary markedly ranging from $19 to $217 million. The IIASA values are the most favorable (lowest) in 5 of the 6 countries, but the least favorable in Iran (though IIASA provides averages across the Middle East, which may affect that result). The IEA values are typically the least favorable with the US EPA values in the middle, except for Russia and Kazakhstan where the EPA values are the highest. Averaging across the three analyses, the largest total benefits (a function of costs and emissions magnitude) appear to lie in Turkmenistan, with net savings of ~$200 million, followed by Russia and the US, with net savings of ~$100 million each.

We also evaluate societal costs when accounting for the monetized environmental impacts. We incorporate the recently described valuation from the Global Methane Assessment (UNEP/CCAC, 2021) that assigns a value of $4400 per tonne methane accounting for the manifold impacts of methane on climate and surface ozone, both of which affect human health (mortality and morbidity), labor productivity, crop yields, and other climate-related impacts. Including those impacts, controlling high emitters in the six countries highlighted here leads to robust net benefits of ~$6 billion for Turkmenistan, ~$4 billion for Russia, ~$1.6 billion for the US, ~$1.2 billion for Iran, and ~$400 million each for Kazakhstan and Algeria. The range across the three mitigation cost analyses is small in this case at ~10% (Fig. 3b).

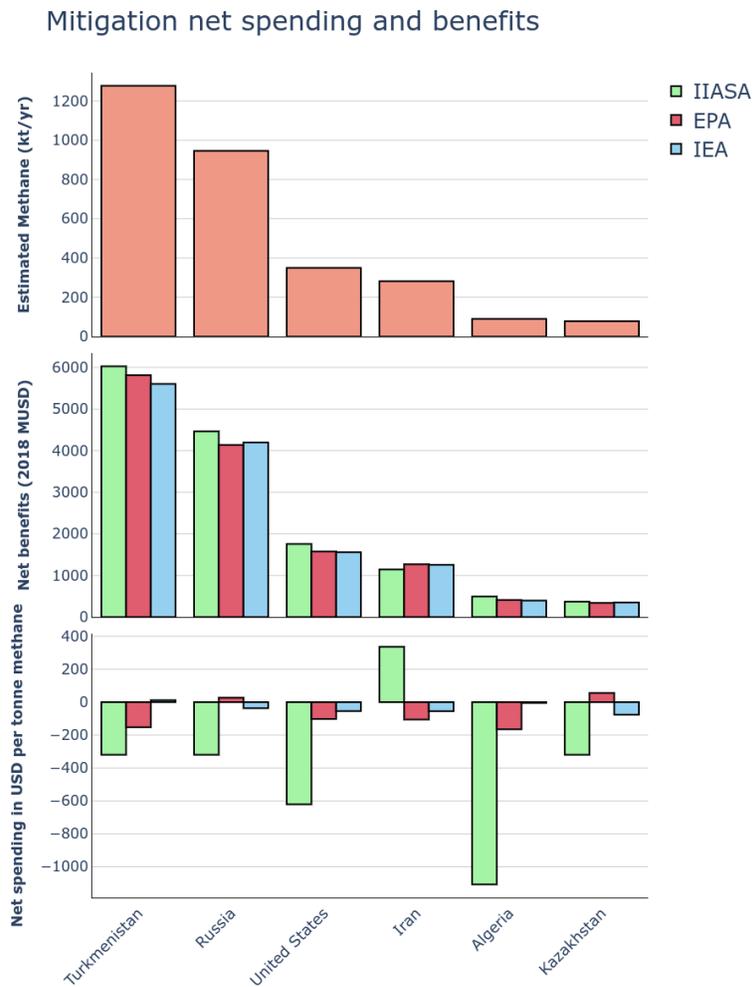

**Figure 3**: Estimated mitigation costs per tonne for high emissions in the oil and gas sector based on the indicated cost analyses (a) and net societal benefits of mitigation of high emitters including monetized environmental impacts (b).

**Discussion**

Based on the power-law distribution of emitters, we derived a detection threshold of 25 $tCH_4.h^{-1}$, in agreement with Varon et al. (2019) using a cross-sectional flux approach to estimate the leakage rates of a major leak in Turkmenistan. For lower emission rates, the number of emitters invisible to TROPOMI far surpasses visible ultra-emitters as suggested by airborne surveys over the Central Valley in California, the Four Corners region, and the Permian basin in Texas (Duren et al., 2019; Frankenberg et al., 2016; Cusworth et al., 2021). High resolution satellite imagery from Sentinel-2 (Varon et al., 2021) or from PRISMA and GHGSat (Cusworth et al., 2021) depict turbulent $XCH_4$ plume structures enabling facility attribution and quantification of leaks above 50 $ktCH_4.yr^{-1}$. These imagers offer limited coverage (tasking mode over small regions) which suggests a combined use with TROPOMI is necessary to achieve monitoring needs. Additional satellite instruments are planned to launch in the near future (e.g., EnMAP, Carbon Mapper, MethaneSAT, SBG, EMIT, CHIME) offering high-resolution images (30m resolution) over selected high-priority areas, precursors to full constellations of imagers covering the globe daily. Until then, and given the robust power-law distribution of $CH_4$ ultra-emitters, the link between intermittent high-resolution imagery and regular low-resolution images from TROPOMI can help fill the gap in coverage. Attribution to specific facilities or operations remains critical to support the development of robust national emissions inventory as defined by the United Nations Framework Convention on Climate Change (UNFCCC), to inform gas operators of accidental releases, and to help regulators on progress in $CH_4$ emission trends.

## Data access

ESRI. "World Imagery" [basemap]. Scale ~1:591M to ~1:72k. "World Imagery Map". April 2021.

OGI <http://www.oilandgasinfrastructure.com/home>

COAL <https://globalenergymonitor.org/projects/global-coal-mine-tracker/tracker-map/>

# Materials and Methods:

The Supplementary Information is structured as follows:
1. Details on the TROPOMI data used in this study
2. Supplementary methods for plume detection, flow rate quantification, and country-level ultra-emitters estimates
3. Uncertainty analysis and measures validation
4. Plumes dataset

These sections include supplementary figures S1 to S16.

## 1. TROPOMI data

### 1.1 General information

We use total column CH4 bias corrected measurements (XCH4 bias corrected) from the spaceborne Tropospheric Monitoring Instrument (TROPOMI). TROPOMI is in polar sun‑synchronous orbit and provides global mapping of atmospheric methane columns on daily overpasses at about 13:30 local solar time with 7 x 7 km nadir pixel resolution (7 x 5.5 km since June 2019). The mission performance report for Sentinel-5 Precursor Level 2 Methane product (ESA 2020) states that the average bias for the comparison against 22 TCCON (Total Carbon Column Observing Network) sites is -0.8% and -0.31% for the standard and bias corrected XCH4 product respectively.

Sentinel-5P data products are released in the netCDF format and the footprints have an irregular geometry. For ease-of-use reasons when applying computer vision algorithms and matching Sentinel-5P observation with HYSPLIT simulations, Sentinel-5P images are reprojected on a regular geometry using the GDAL library prior to any other processing (GDAL, 2021).

The XCH4 bias corrected is a Level 2 data product released by the European Space Agency (ESA), expressed in parts per billion (ppb), derived from the Level 1 data product (radiance and irradiance measurements). In our analysis, we do not use Level 1 data and only rely on Level 2 data. However, we also use the Level 2 data quality (qa_value) product. To ensure robustness in our results, we exclusively take into account pixels for which qa_value > 75.

Our analysis is based on data sensed over two full years between the 1st of January, 2019 to the 31th of December, 2020, extracted continuously 2 to 5 days after sensing.

### 1.2. Sentinel-5 Precursor observations availability

Sentinel-5 Precursor has a daily revisit time, but observations are incomplete. For various reasons (clouds, humidity, albedo, etc) a significant fraction of the pixels are missing (see figure S1). On average in 2019, on a 0.05×0.05 degree regular grid, S5P successfully retrieved a XCH4 measure for 7% of daily onshore pixels. The distribution of missing pixels is not homogeneous however, as some places (e.g. equatorial zones) are essentially missing whereas some drier places have more than 100 measures per year. Considering only onshore pixels with at least 10 valid XCH4 measures in 2019, the daily proportion of covered pixels increases to 13%. TROPOMI does not provide any reliable measure offshore at this time.

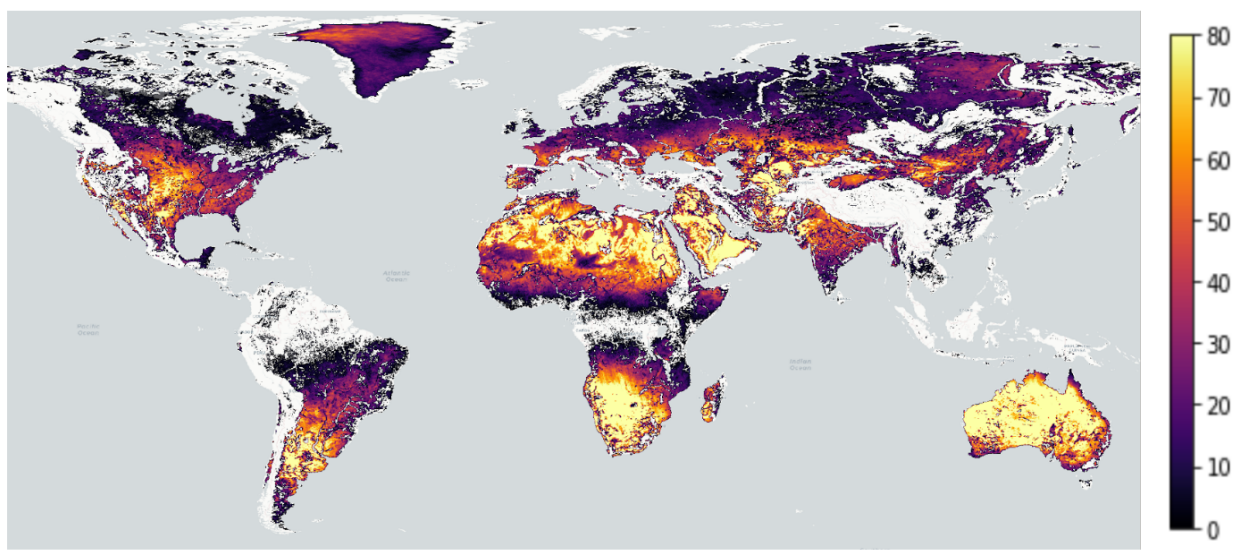

**Figure S1**: Sentinel-5P coverage for Level 2 XCH4 data product in 2020. The value of each pixel corresponds to the number of days for which Sentinel-5P provided at least one valid (after quality filtering,[ESA, 2020]) measurement, for the corresponding area during year 2019. Note that 80 is a hard threshold set for clarity; some pixels exceed this value.

## 2. Supplementary methods for plume detection, flow rate quantification, and country-level ultra-emitters estimates

The general framework used here is the following:
    1) detect ultra emitters using an automated algorithm and human labeling
    2) quantify their flow rate using Forward Concentration simulations,
    3) aggregate and adjust emissions for coverage and leak duration,
    4) perform a country-scale cost/benefit analysis.

We now describe the procedure and evaluate each step including associated uncertainties.

### 2.1. Plume detection

#### 2.1.1. Background estimation and plume detection algorithm

At every orbit, Sentinel-5P produces 13 to 14 images (or tiles) from the South Pole to the North Pole with a 2600km swath width. Each tile is processed with a plume detection procedure as follows.

1. The image is first denoised using Gaussian filters (Buades *et al.*, 2010).
2. Local standard deviation and background values are computed dynamically as follows. In the literature, background methane on S5P images is estimated by either taking the value of the pixel in the vicinity of a detected plume in the upwind direction or by taking the median of the image (Pandey *et al.*, 2019, Varon *et al.*, 2019). As we want to estimate background before identifying methane plume, we cannot apply the first method. The second is also a poor match in this case, as we process large tiles on which methane background is not homogeneous. Here instead, for each pixel, we consider the 11 by 11 pixels patch centered around it and compute standard deviation at this pixel as the standard deviation of the patch. The background value at this pixel is computed as

$$median \text{ if } \frac{mean - median}{std} > 0.3$$

$$l \times median - (l - 1) \times mean \text{ otherwise}$$

where $median$, $mean$ and $std$ denote respectively the median, mean and standard deviation of the patch. This method is commonly used for robust background estimation in noisy astronomical images analysis (cite Astropy, A. M. Price-Whelan *et al.*, 2018). The background value is computed as $l \times median - (l - 1) \times mean$ to be robust to the influence of plume pixels in background estimates, where $l$ is typically equal to 2.5 (cf. Section 2.1.2). If the pixel distribution is strongly skewed, the difference between the mean and the median would have a significant impact on the background estimate, which might introduce a bias in our background estimate. Thereby, if the condition $\frac{mean - median}{std} > 0.3$ holds, the background is the median of the patch.

3. Plumes are then segmented. An anomaly map is defined as

$$AnomalyMap = Image - Background - k \times StandardDeviation$$

where $Background$ and $StandardDeviation$ maps refer to those computed at step 2. On this anomaly map, contiguous groups of positive pixels are selected as plume candidates, setting $k = 3$.

4. Contiguous but distinct plumes (i.e. 2 or more plumes that are emitted by distinct source but whose footprints overlap) are then separated (see figure S2). A sharpening kernel is applied to the whole background-corrected denoised image to tackle the edge vanishing issue implied by Gaussian denoising (Buades et al., 2010), and contiguous plumes are separated using watershed segmentation (Beucher *et al.*, 2000).

5. Any detected plume is discarded if the average of the XCH4 enhancement of the pixels in the plume is below $avgenhancement$ **or** the number of pixels with a QA higher than 75 is below $minqapixels$. We typically use $minpixels = 5$ and $avgenhancement = 25$.

6. For all plumes that have not been discarded at step 5, a first estimation of the source location is obtained by following the upwind direction from the centroid of the plume. The last pixel found within the plume polygon is then chosen as the source location. This source location estimate is then going to be refined by human labelling (see section 2.2. Flow rate quantification).

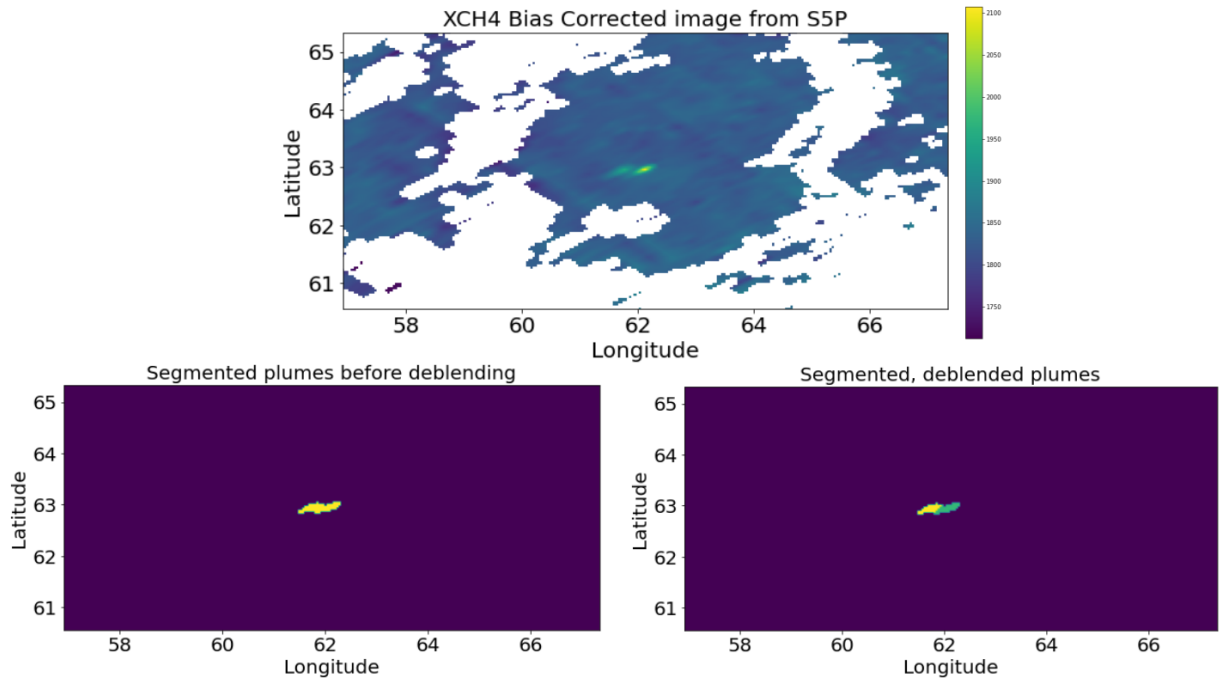

**Figure S2**: major steps of the detection algorithm. The two methane plumes visible on the XCH4 image (top) originate from two nearby sources on a Russian pipeline (probably routine maintenance where leaks come in pairs). The methane anomaly detection output (bottom left) is a contiguous set of pixels. After the deblending step, the algorithm retrieves two contiguous but distinct plumes (bottom right).

### 2.1.2. Parameters estimates

The algorithm includes several predefined parameters used in the Gaussian denoising filter (kernel size and standard deviation) and the sharpening filter (intensity of the central pixel of the kernel with respect to its neighbors) that must be optimized, as well as the parameters described above (cf. section 2.1.1): $k$, $l$, $minqapixels$, $minpixels$, and $avgenhancement$. These parameters have been set such that the algorithm successfully retrieves some relatively well-known methane emissions, including leaks in Turkmenistan [Varon et al., 2019], or confirmed events (without official quantification) in the vicinity of Hassi Messaoud oilfield in Algeria [Private Communication, Sonatrach]; and along Russian pipelines (see figure S2)). The set of parameters has also been defined to limit the number of false positives (around 95% accepted) when labelling the detections manually. This rate is sufficiently large so that new plumes with lower flow rates have been discovered, while controlling the number of false positives.

### 2.1.3. Individual plume labelling

All plume candidates identified at step 6 of the algorithmic procedure are submitted to a human labeler. The human labeler looks for evidence that the candidate plume is a false positive detection (hence should be rejected) according to the following criteria:

- The plume direction is inconsistent with the wind direction from the ECMWF-ERA5 reanalysis product (100m u- and v-wind components) (Copernicus Climate Change Service, 2017). The plume is discarded if its direction diverges from the wind direction at the round hour before sensing. Figure S3 illustrates the empirical angles distribution for both accepted and discarded plumes; it highlights that there is *a posteriori* an empirical acceptance threshold around 30 degrees (above which unambiguous methane plumes are still accepted).
- The plume spatial distribution correlates with spatial gradients in the Surface Albedo SWIR product provided by Sentinel-5P. Biases induced by the albedo in the XCH4 retrievals from Sentinel-5P are well-known but not properly removed in the official L2 product (ESA 2020). We discarded all the detected plumes with a strong correlation with the surface albedo to avoid false positives (Fig. S4).
- Similar to the correlation with surface albedo, we removed from our analysis all plume candidates matching spatial patterns visible in optical images (ESRI World Imagery). The rationale behind this removal is the same as for the previous item (Fig. S4).

At this stage, the labelling includes the attribution of the detection to an activity sector, or is labelled "Other human activity" for undefined plume origins. This category can be either "Oil and Gas", "Coal", or "Other human activity". This decision is based on the knowledge of methane-emitting activities on the ground, derived from geospatial data sources such as Oil and Gas Infrastructure and Petrodata v1.2 "Other human activity" refers to methane emissions from complex areas where multiple source candidates are present (i.e. large metropolitan areas) or when geospatial data includes no potential known source of $CH_4$. Large metropolitan areas where large anomalies were detected, such as Karachi, Lahore, Delhi, or Dhaka, often include landfills and waste management facilities, large natural gas city networks, or coal stockpiles that could all emit large amounts of $CH_4$.

Figure S5 illustrates various plumes detected around the world by the algorithm and validated by the human labeller.

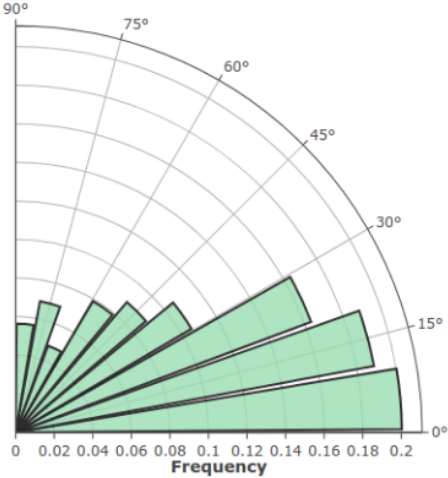 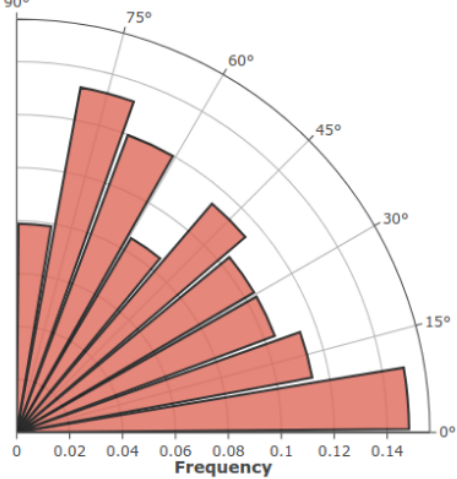

**Figure S3**: Distribution of the angles between methane plumes direction and ERA5 100m wind direction, for plumes accepted (left) and rejected (right) by the human labeller. The direction of the detected plumes is computed as the first principal component in the singular value decomposition of the vertices of the plume polygon. Note that false positive plumes may have been rejected either for wind direction or for e.g. albedo pattern matching. The false positives histogram is based on a random sample of 500 false positive plumes.

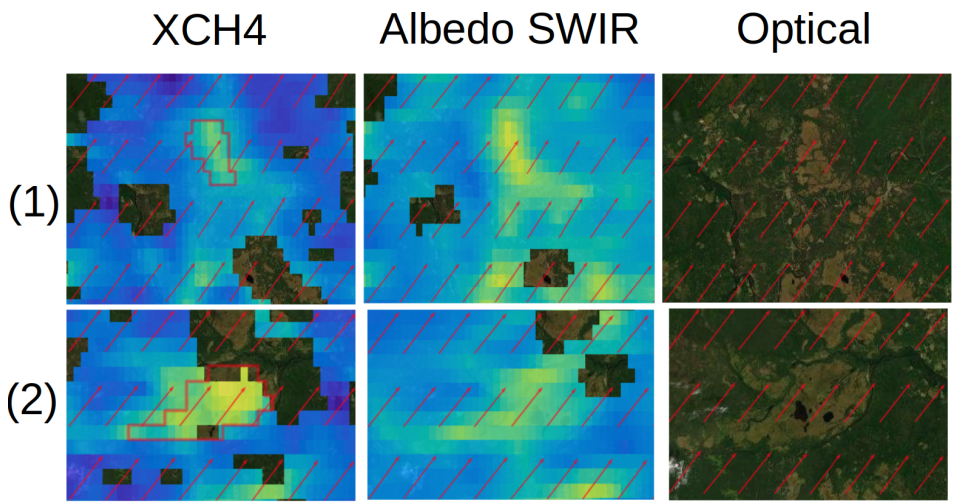

**Figure S4**: examples of false positive detections discarded by the human labeler. Sentinel-5P XCH4 bias corrected images(left column); corresponding S5P SWIR albedo images (middle column); optical images (right column). On all images, red arrows represent the wind data. In row (1), the pattern detected on the XCH4 image (red polygon) is also visible in the albedo SWIR image and on the optical image. In addition, the wind direction does not match the direction of the detection: this detection must be discarded. Likewise, in row (2), the detected pattern is visible on both albedo SWIR and optical image. Even if the wind direction matches the major axis of the detected pattern, it must be discarded. Image credit: ESRI.

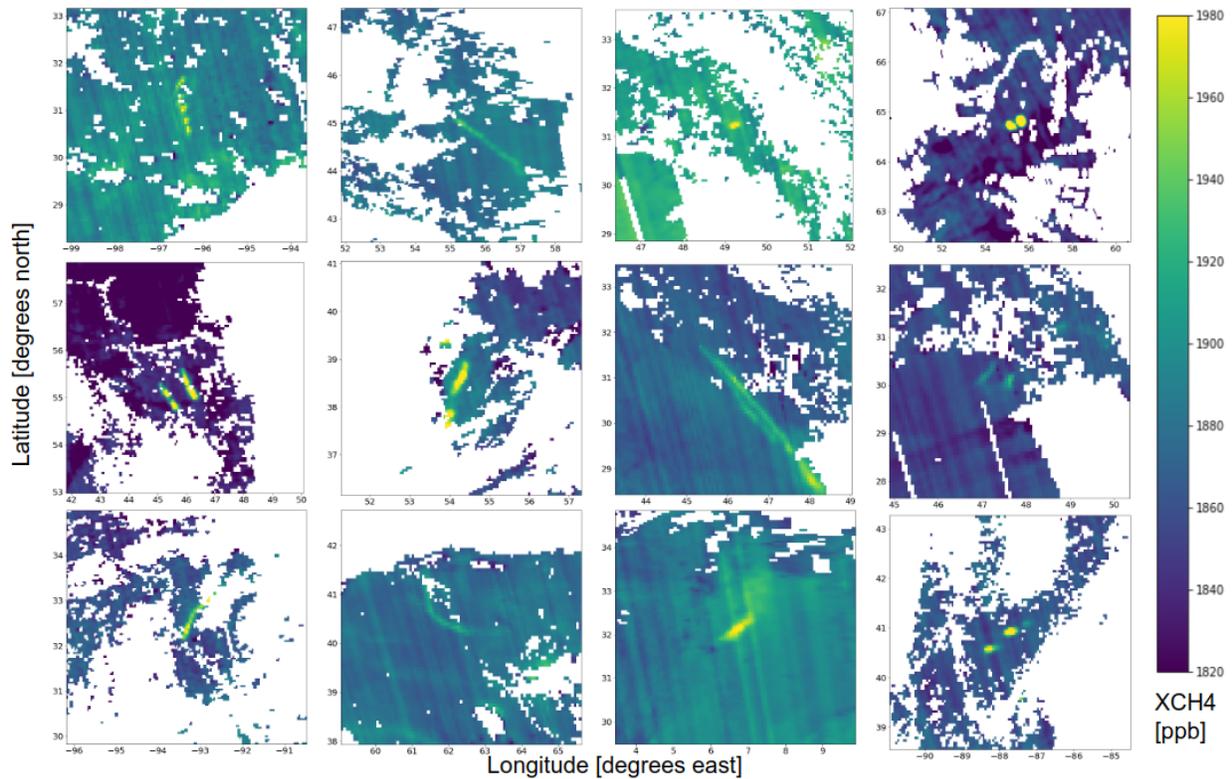

**Figure S5**: Examples of detected plumes validated by human labelling in various tiles from the L2 XCH4 TROPOMI retrievals. Clouds and ocean pixels are shown in white. For readability, all available pixels are shown here, without applying the qa_value filter.

## 2.2. Plume modeling and flow rate quantification

This step aims at quantifying the emission flow rate of all the plumes that have been detected by the algorithm and validated by the human labeller. The methodology is similar to the mass balance approach from Pandey *et al.* (2019).

### 2.2.1. Atmospheric modeling

For each detected plume, we simulated the observed enhancement using the Lagrangian particle dispersion model HYSPLIT (A. F. Stein *et al.*, 2015) in forward mode. We run the HYSPLIT model in concentration mode on a 0.01×0.01 degree grid, significantly higher than the resolution of Sentinel-5P. The particles representing an air mass containing a fixed amount of $CH_4$ are released continuously assuming a wind-following Gaussian puff in the horizontal, with particles mixing vertically over the prescribed Planetary Boundary Layer (provided by the meteorological input fields). The number of elements released at each hourly cycle is 2500. Assuming that the observed plumes are in steady state, the start of release is set 7 hours before sensing time which is sufficient to model the visible enhancements for 67% of the detections. If the observed plume extends beyond the simulated plume, new simulations are performed with earlier release times until the plume length matches the observed one. The particles are released at 10 meters above ground level to account for high-pressure injection heights. The meteorological data used for the HYSPLIT simulations come from the Global Forecast System (GFS) by the National Centers for Environmental Prediction (NCEP) at 0.25-degree and hourly resolutions. When GFS is not available on the NOAA FTP server, we use the Global Data Assimilation System (GDAS) meteorological data from NCEP at 1-degree and hourly resolutions. The model simulates plumes originating from the source location estimated at the previous section (cf. section 2.1.1. step 6.). Simulated plumes are reprojected on the observed Sentinel-5P geometry.

### 2.2.2. Flow rate quantification

A mask is formed from HYSPLIT plumes by selecting all methane-enhanced pixels in the simulated plume whose enhancement is bigger than 10% of the most intense pixel enhancement (i.e. removing the edges of the plume represented by too few particles). Observed Sentinel-5P enhancements are calculated as the difference between XCH4 values and background (cf. section 2.1.1). The emission rate Q is then quantified by comparing TROPOMI-observed and HYSPLIT-simulated XCH4 enhancement restricted to the area described by the HYSPLIT mask, projected on Sentinel-5P's geometry, with

$$Q = Q_\omega X / X_\omega$$

where, X and $X_\omega$ are the XCH4 enhancements (in parts per billion) for TROPOMI and HYSPLIT plumes respectively, and $Q_\omega$ is the constant emission rate used in the HYSPLIT simulation. Several factors bring uncertainty to the estimated flow rate $Q$. Refer to Section 3.1. of the Supplementary Information for an analysis of the uncertainty of the estimated flow rates.

Similar to the detection stage, quantification results are manually checked by a human labeler. In particular, we discard false positives when the simulated plume direction diverges significantly (*a posteriori,* the empirical threshold is 30 degrees, see figure S6) from the observed plume direction. Wind direction mismatch indicates that the GFS or GDAS weather data is not consistent with the observed plume direction. Figure S6 quantifies the angle between simulation and observation when the quantification is rejected for direction divergence. Another option for the human labeller is to state that the flow rate of detection is impossible to quantify. This can be due to a multi-source environment for which our method is not suited, or a small wind velocity (i.e. a compact plume with no well-defined direction) setup in which quantification methods do not apply (Varon *et al.*, 2018). For a limited number of detected plumes, an ensemble of HYSPLIT simulations were performed using different simulation durations and source locations to improve the fit between observed and simulated plumes, evaluated following the same steps as described above. Figure S7 shows HYSPLIT for both accepted (top and middle rows) and rejected (bottom row) flow rate quantifications. In summary, 518 plume quantifications have been rejected and 702 accepted out of 1,220 detections related to oil and gas during the timeframe of our study (2019-2020).

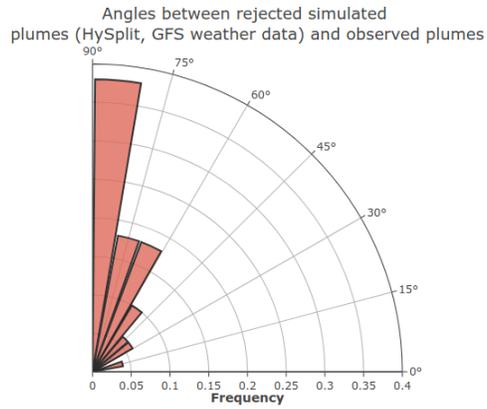

**Figure S6**: Angles between detected and simulated plumes, in the case where flow rate quantification is rejected because of a mismatch between detected and simulated plume directions. Most of the plumes in this case form an angle with simulated plumes that is bigger than 30 degrees.

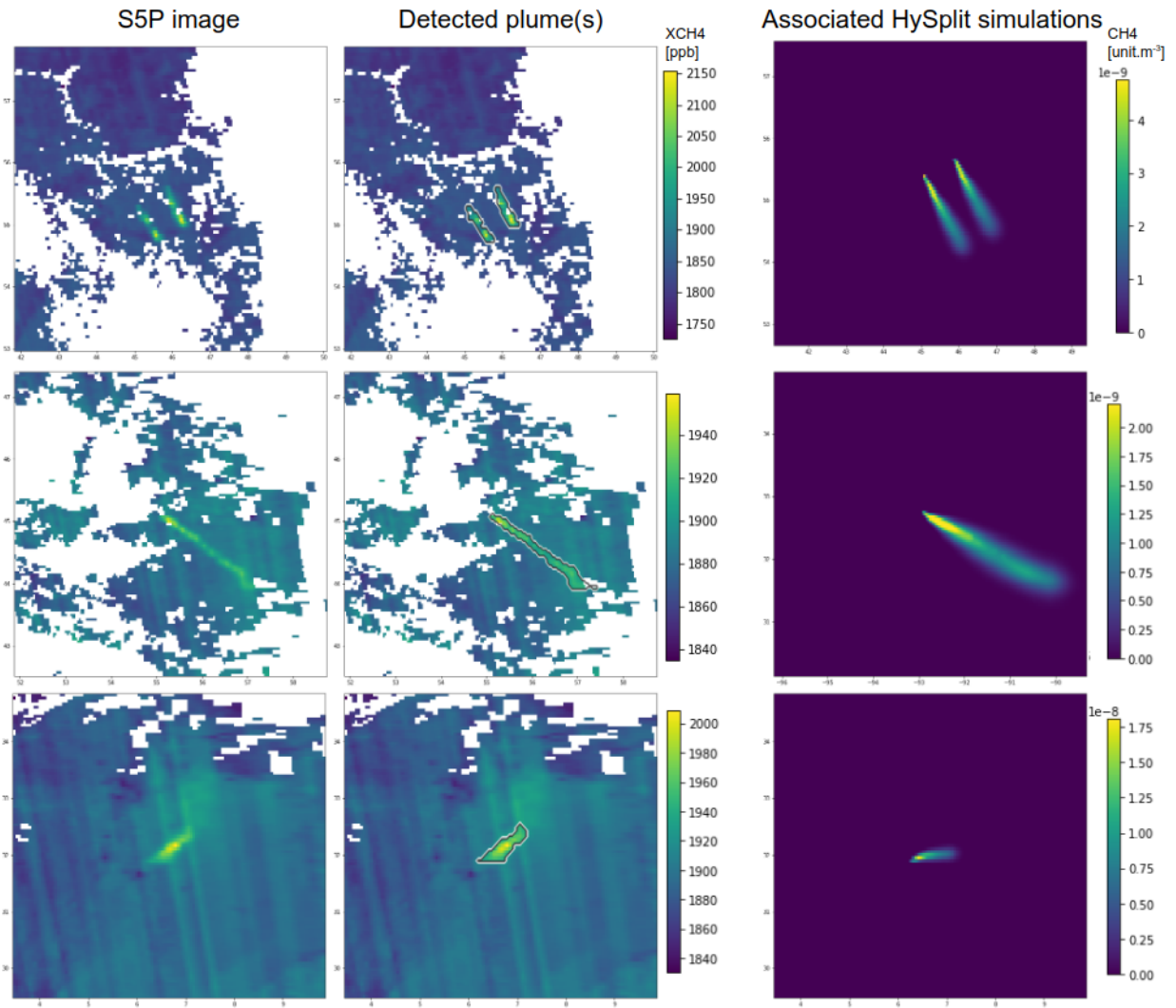

**Figure S7**: TROPOMI images (left column), plumes detection overlayed on TROPOMI images (middle column), associated HYSPLIT simulations (right column). On top and middle rows, simulated plumes lengths and directions match the observed plumes; these quantifications have been accepted by the human quantifications checking. On the bottom row, there is a mismatch between observed and simulated plume directions; this quantification is rejected by human checking. On the top row, two plumes are shown on the same simulation for completeness, but they are handled independently in the quantification algorithm.

### 2.3. Country-Level Ultra-Emitters Aggregation

From detections and quantifications, we derive aggregated figures to estimate methane emissions from ultra-emitters at national scale. Three key figures are provided for each area of interest and time period in addition to the leak duration for each observed plume:
1. **Observed emissions**, which are the sum of emissions due to detected leaks.
2. **Coverage,** i.e. the number of actual measurements during the selected period, of sufficient quality to detect a methane plume. This quantity is a positive floating number with a maximum equal to the number of days over the observing period. Details on this metric are given in section 2.3.1 below.
3. **Leak duration**, i.e. the actual duration of any observed events. Three scenarios are presented (cf. section 2.3.4) to account for the full duration of any detection based on continuity (leaks visible on consecutive images) and length of observed plumes.
4. **Estimated emissions,** i.e. an estimate of the emissions that would have been observed given perfect coverage. Details on how we adjust for coverage are given in section 2.3.2 below.

### 2.3.1. Coverage

Coverage quantifies the number of valid readings provided by Sentinel-5P during a selected time interval. We compute coverage indicators by splitting each region into elementary patches. On each patch, a logistic regression model detailed below predicts if it would have been possible to detect a methane plume given atmospheric conditions and quality assurance data (the patch is then marked as "valid"). The ratio of valid patches over all patches for a given day represents daily coverage. Daily coverage is then aggregated by adding up daily coverages into monthly, quarterly and yearly coverage numbers. The coverage for a given period is a number between zero and the number of days in the period.

Estimating emissions due to ultra-emitters in an area of interest (AOI) requires estimating "coverage", i.e. quantifying the number of days for which ultra-emitters could be detected in the area using Sentinel-5P images. To compute this number, we use the following algorithm.
1. Split the AOI into patches. The dimensions of each patch is 120*120km. Each patch overlaps with half of its right, left, top and bottom neighbors, to ensure that a pixel that is at the edge of some patch is also at the center of another patch.

2. For each patch, apply a logistic regression model whose output is 1 if the quality of the patch is good enough for the detection algorithm to detect a methane plume, 0 otherwise. Details on the training of this logistic regression model are given below.
3. For a given day and a given AOI, the coverage is defined as a floating number equal to the number of valid patches divided by the total number of patches in the area of interest.
4. For a given period, the coverage is the sum of daily coverage for the period.

To define the dimensions of the patches, we plot the distribution of the length of the detected plumes (figure S8). As the 80% quantile of this distribution is 60km, this means using 120x120km patches ensures that most plumes are entirely included in at least one patch.

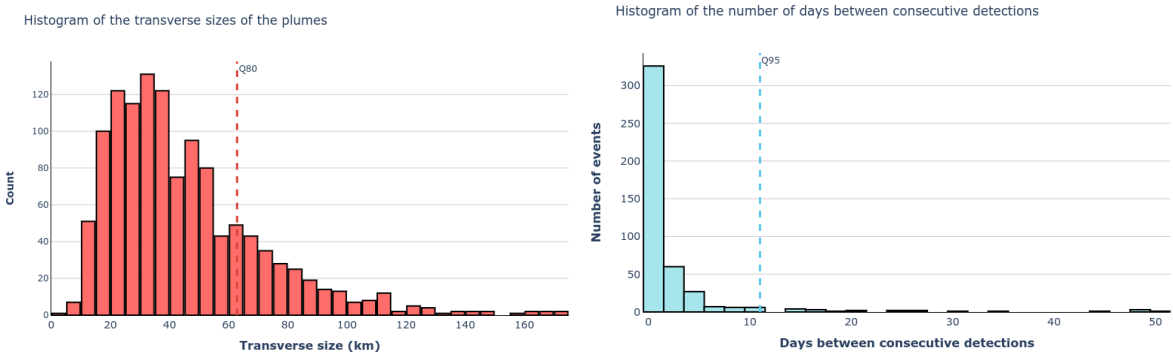

**Figure S8**: Histogram of the length of the detected plumes (left). Histogram of the number of days between two consecutive detections in the same patch (right); 14 days corresponds to the end of the fat tail of this histogram and is above the 95th percentile.

To train the logistic regression model mentioned in the preceding paragraph, we first build a dataset of positive and negative observations based on the image mask (i.e. missing pixels due to weather, albedo, etc.), using the following process for each detected methane plume. Note that the input of the logistic regression model is not the $XCH_4$ pixel values, but the distribution of the QA values of the pixels. To build this dataset, we use a subset of 300 detected methane plumes, and apply the following process:

1. Crop a 120*120km patch containing a detected methane plume.
2. Downsample the patch image using a mask sampled at another random location in an S5P image.
3. If the plume detection algorithm still detects the methane plume, the mask is given label 1, otherwise 0.
4. The process is repeated until we obtain a balanced dataset with 10,000 observations.

Logistic regression is then trained on this dataset to discriminate between valid (label 1) and invalid (label 0) patches. We then apply this model daily on each patch to determine if detection is possible or not on each particular date and patch. These classification results are then aggregated into our measure of coverage.

## 2.3.2. Observed and Estimated Emissions

We estimate total emissions by scaling observed emissions as follows.

**Adjusting for coverage loss:**

To adjust for coverage, for each AOI over each time period, we first compute the number $n_{obs}$ of observed emission events, and the coverage $c$ described in section 2.3.1 above, i.e. the number of days for which S5P images were complete enough for ultra-emitters to be detected.. We then estimate the total number of emission events over the period as

$$\frac{n_{days}}{c} n_{obs}$$

by scaling the number of observed events, where $n_{days}$ is the number of days and $c$ is the coverage in the period. Total emissions for the period are then estimated from observed emissions using the same

$$\frac{n_{days}}{c}$$

scaling factor. This implicitly assumes that emission events and rates are independent from weather patterns over the period, which might not be true in Russia in winter. If leaks are more common, it would result in an under-estimatation.

**Quantifying uncertainty due to coverage.**

We use a negative binomial model to quantify the uncertainty introduced by these adjustments for coverage. This can be traced as far back as the work of (Student 1907), see also (Hilbe 2011) for a more recent discussion. Each area of interest and time period is treated independently in the following way.

1. Compute coverage $c$ for the AOI during the given period.
2. Estimate the number of leaks that would have been detected given full coverage during this period, as

$$n_{est} \sim NB(n_{obs}, p)$$

Here $n_{est}$ is the estimated number of leaks, $n_{obs}$ is the observed number of leaks, $p = c/n_{days}$, where $n_{days}$ is the number of days and $c$ is the coverage in the period, and $NB$ stands for the negative binomial probability distribution. For a given number of observed events $n_{obs}$ detected in a fraction $p$ of all the observations, $NB(n_{obs}, p)$ is the distribution of the number of events that would have been detected in the full period $n_{days}$ assuming emission events are independent identically distributed Bernoulli random variables with probability $p$. The mean of this probability distribution is $\mu = \frac{n_{obs}}{p}$, and its variance is $\sigma^2 = \frac{n_{obs}(1-p)}{p^2}$. Note that while the mean $\mu$ of this distribution matches the estimated total number of emission events used in the previous paragraph, this model allows us to produce confidence bounds and show 90% symmetric confidence intervals.
3. Estimate the distribution of total emissions in the AOI after adjusting for coverage. The aim here is to estimate the distribution of total emissions from observed **and** non observed sources. This distribution is sampled as follows.
    - Pick an estimated number of leaks: $n_{est} \sim NB(n_{obs}, p)$
    - For $i \in \{1,..., n_{obs}\}$, take the $i^{th}$ quantified detection among those observed in the AOI during the period, write its rate as $q$, and sample an emission rate $r_i \sim N(q, q \times 0.45 / 1.96)$. The rationale for this choice is that the median relative uncertainty on the estimation of emission rates is 45% (cf. SI section 3.1.).
    - For $i \in \{n_{obs} + 1,..., n_{est}\}$, randomly pick a quantified detection among those in the AOI during the period, write its rate as $q$, and sample an emission rate $r_i \sim N(q, q \times 0.45 / 1.96)$.
    - Sample total methane emissions as

    $$E = \sum_{i=1}^{n_{est}} r_i \times H_{emit, i}$$

    where $H_{emit, i}$ is the estimated duration of emission $i$ (which depends on the duration scenario; cf. section 2.3.4).
    - Repeat $N$ times to sample the distribution of total emissions, and compute Monte Carlo sampling confidence bounds.

Because the mean µ of the negative binomial distribution matches the estimated total number of emission events, and because emission rates are sampled independently, the sample mean of total emissions obtained using this procedure converges to the scaled total computed from observed emissions in the previous paragraph, given enough samples. The sampling approach however allows us to compute confidence intervals on coverage adjusted emissions.

### 2.3.3. Leak duration scenarios

As the satellite revisit time is about 24 hours (except for places near the equator and polar regions), the exact duration of each detected emission event is unknown so we build three scenarios to estimate emission duration $H_{emit}$.

- The first scenario is a "*lower bound*". Each quantified detection is matched with a HYSPLIT simulation with a duration $H_{sim}$ ranging between 2 and 10 hours. In this lower bound scenario, where we assume that emissions only started on the same day, and set $H_{emit} = \gamma \times H_{sim}$. In a simple model where satellite overpass is at noon, emission start time is uniformly distributed over 24h and methane remains above the detection threshold for nine hours, true emission duration is proportional to 2.1 times observed plume duration, so we set $\gamma = 2.1$ here.

- The second scenario is the "*intermittency estimation*" framework. It takes into account the fact that some emissions last for a few hours while others last for several days. For each detected event, based on data availability and detections in its neighborhood in the previous and next days, we assess whether the emission lasted for several days or not (details on the methodology are given below). In the first case, we set $H_{emit} = 24h$, in the second one, $H_{emit} = \gamma \times H_{sim}$. We do not define leak durations beyond 24 hours because the adjustment for loss of coverage compensates for days without observations, hence compensating for leaks lasting several days.
- The third scenario assumes "*continuous emission*", meaning that each detected source is supposed to last 24 hours. Although this scenario represents an "upper bound" to the duration of an individual leak, the satellite is likely to miss intermittent emissions outside of overpass time, which will bias our observations downwards.

The *intermittent scenario* represents the most physically plausible scenario of the three and is therefore selected as our reference in the comparison to the inventories and the cost-gain analysis. To clarify how we applied it to estimate the emission duration, we describe here the process for each detected plume in more details:

1. Find all the patches intersecting the plume footprint (the definition of the patch is the same as in SI 2.3.3.)
2. Find the nearest date in the past and the nearest date in the future for which at least one of these patches is valid (the definition of a valid patch uses the same logistic regression model as in the coverage definition). We set a hard threshold to 14 days: if there is no valid patch 14 days before and after the plume detection, then the plume is considered as *intermittent*. The choice of 14 days is shown in figure S8 (right panel), 14 days corresponding to the end of the fat tail of the histogram.
3. If either the next or the previous valid patch contains at least one detection, then the plume is considered as a *continuous* leak. Otherwise, it is considered *intermittent*.
4. We take $H_{emit} = 24h$ for continuous emissions and $H_{emit} = \gamma \times H_{sim}$ for intermittent emissions.

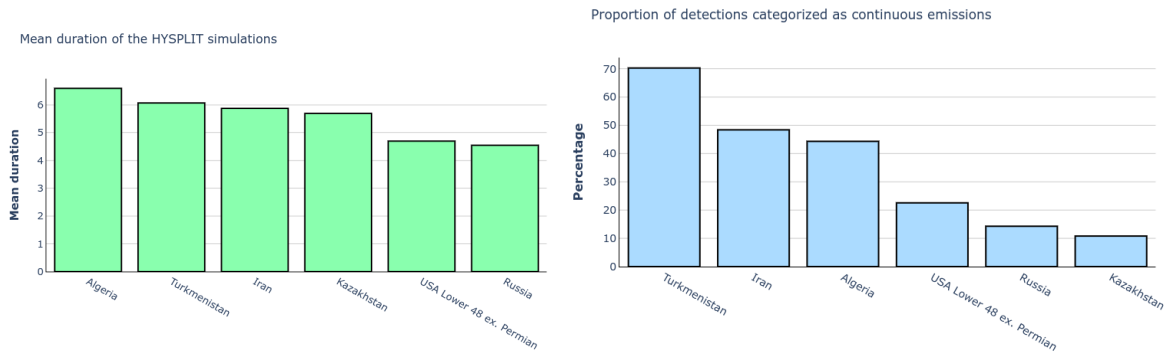

**Figure S9**: mean release duration in the HYSPLIT simulation associated with the flow rate estimates in each country (left); percentage of plumes categorized as "continuous" in the *intermittent scenario* in each country (right).

Countries with the most continuous plumes are also those in which the release durations in the HYSPLIT simulations are longer.

### 2.3.4. Validation of the coverage loss

The adjustment for the loss of coverage depends on the sampling rate for a given country. To evaluate the robustness of our estimated emissions when only a limited number of detections is available (e.g. over Russia or Iran), we performed the following experiment: we subsampled S5P images from one of the countries with the most complete observation set (e.g. Turkmenistan) to match the number of observations from one of the countries with the lowest coverage (e.g. Iran). By repeating this subsampling procedure, we can estimate the error due to a low number of detections, and in parallel, evaluate our uncertainty estimate. Following this procedure, we estimated the emissions from Turkmenistan (where coverage is high - around 118 in yearly average) by subsampling the available images. We randomly censored observations until the yearly mean coverage reaches 22 which corresponds to the coverage over Iran, the smallest among the studied countries. We then apply the aggregation algorithm detailed in SI 2.3.2 to the censored data to calculate the estimated emissions, and we repeated the process 100 times to produce a statistical distribution of the subsampling. The results are shown on figure 2.e. in the main text. The mean of the 100 estimates based on censored data for Turkmenistan is 1.26Mt (associated 90% confidence interval: 0.87Mt to 1.64Mt) whereas the estimate based on full data is 1.28Mt (90% confidence interval: 1.15Mt to 1.41Mt). Furthermore, the dispersion of the estimated emissions based on subsampled data fits the associated confidence intervals.

### 2.3.5. Areas used for country-level emissions estimation

In the USA, the Permian basin contains a large number of methane anomalies which are detected by our algorithm. These detections consist of multiple overlapping plumes from numerous small to medium sources, hence not from single emitters. For that reason, we chose to remove the detections over the Permian basin from our analysis (cf. figure S15). All our estimates and comparisons to the national US inventory estimates exclude the Permian, as explained in the main text.

In several countries, we also limited our observed area to the most active zones in terms of O&G production and transmission activities. We excluded the areas with high coverage loss who are very unlikely to contain large oil and gas related to methane leaks because they neither contain major midstream nor upstream infrastructures, and might introduce a negative bias when their coverage is very low (over-estimation of data loss; for example is Russia, the excluded area has a rate of valid measures 50% smaller that the areas taken into account in 2020, see figure S1). For these reasons, we chose to remove sub-regions from the polygons used in our analysis, in Russia, Kazakhstan, Iran and Algeria (cf. figure S10). The regions we remove in Iran are not major O&G producing areas and have a very low coverage due to rough terrain and mountains; they contain only three detections presumably related to oil and gas activities. The regions we removed in other countries do not contain any detection. The map on figure S10 shows the polygons taken into account in our analysis.

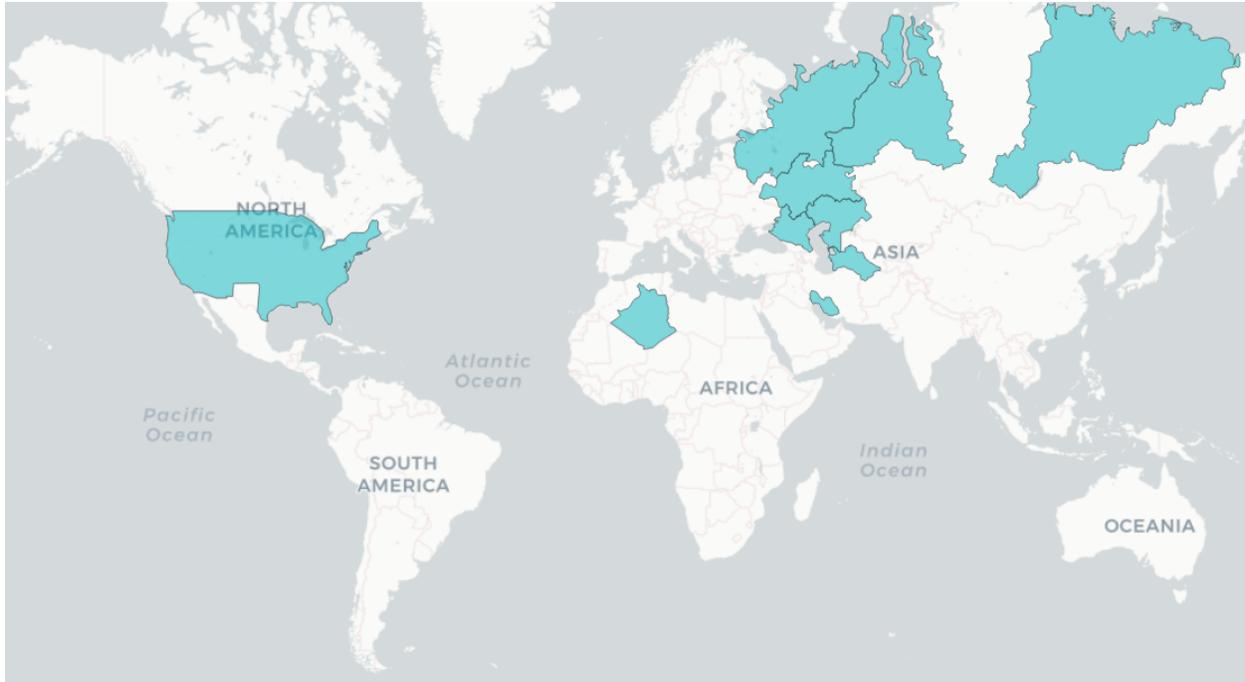

**Figure S10**: polygons taken into account for estimating country-level ultra-emitters methane emissions.

## 3. Uncertainty analysis and measures validation

### 3.1. Analysis of the uncertainty and sensitivity to model parameters

#### 3.1.1. Method

Uncertainty in source rate estimation mainly stems from uncertainty in the model input parameters. We use a methodology similar to Pandey et al. 2019 to estimate the uncertainty of the flow rates we compute. Estimations can vary greatly depending on:
- uncertainty on the Sentinel-5 Precursor measurements,
- errors in meteorological data driving our HYSPLIT simulations,
- uncertain background quantification,
- uncertain longitude and latitude of the source location.

In order to evaluate the magnitude of these variations, we ran a sensitivity analysis on 200 plumes randomly selected among the methane plumes assigned to oil and gas activities we detected in 2019-2020. For each parameter bringing uncertainty to the flow rate estimate, we build an ensemble of simulation with different values for the concerned parameter. The uncertainty associated with the parameter is taken as the standard deviation of the ensemble. For each methane plume detected, input parameters iterate over the following scenarios.

- latitude and longitude with one reprojected Sentinel-5 Precursor pixel variation around the estimated source, to evaluate uncertainty from source location. This leads to a set of 9 flow rate estimates for each plume, whose standard deviation is thereafter noted $\sigma_{location}$
- Two meteorological driver data sources: GFS 0.25 degree, GDAS 1 degree, to represent the transport model uncertainty. The standard deviation of these two measures is noted $\sigma_{weather}$
- Simulation start time offset by ± 2 hours - with an hourly sampling - around the estimated optimal start time (determined by the human labeler), to take into account the influence of the release duration. The standard deviation of the five estimates derived thereby is noted $\sigma_{offset\ hour}$
- Four different background estimation methods are tested - all detailed in the dedicated paragraph below. The standard deviation of these estimates is noted $\sigma_{background}$.
- For each image, the measurement error from TROPOMI is given as a dataset named *methane_mixing_ratio_precision*; which we propagate in our flow rate estimation algorithm to obtain a measure uncertainty $\sigma_{measure}$, (Pandey *et al.*, 2019).

Once we know the uncertainty linked to each parameter, given these parameters are all independent, we can apply the law of propagation of uncertainty (Williams, 2016, Pandey *et al.*, 2019) and compute the combined uncertainty by summing these errors in quadrature

$$\sigma_{total} = \sqrt{\sigma^2_{location} + \sigma^2_{measure} + \sigma^2_{weather} + \sigma^2_{offset\ hour} + \sigma^2_{background}}$$

### 3.1.2. Background Estimation Scenarios

The choice of the method used to compute the background is crucial, since all the estimations we perform are based on methane enhancement, itself linked to the background estimation. In our framework, we let aside the methods in the literature which required manual estimation of the background. This includes for example the choice of a pixel located upwind (Pandey *et al.*, 2019, Varon *et al.*, 2019). Instead we compute the background automatically from the median of the pixels in a bounding box of 1x1 degree around the source locations. The enhancement of the image is then obtained by subtracting the median to the pixel values and setting negative values to 0. This simple method tends to introduce a one-side bias due to the noise in the pixel values. Therefore we derived a second method, where we set to zero the pixels below one standard deviation of the image. This correction is meant to avoid misinterpreting the noise in the S5P image as $CH_4$ concentration variations, which could introduce a negative bias in the emission rates. Analogously, we derive a local version of these two methods, which uses the background estimation method explained at section SI 2.1., to yield more robust estimates in case of partially degraded observations. This leads to four methods to compute the enhancements: median of the neighborhood, the method explained in the section SI 2.1., and a version of these two methods in which the smallest pixels are set to 0.

### 3.1.3. Results

The results of the uncertainty analysis are displayed on figure S11. The median of the total relative uncertainty is 45%. The parameter responsible for the largest uncertainties is the source location (26%). In comparison, background estimation method and error propagated from Sentinel-5P XCH4_precision data product have a limited impact on the uncertainty with relative standard deviations respectively of 10% and 9%.

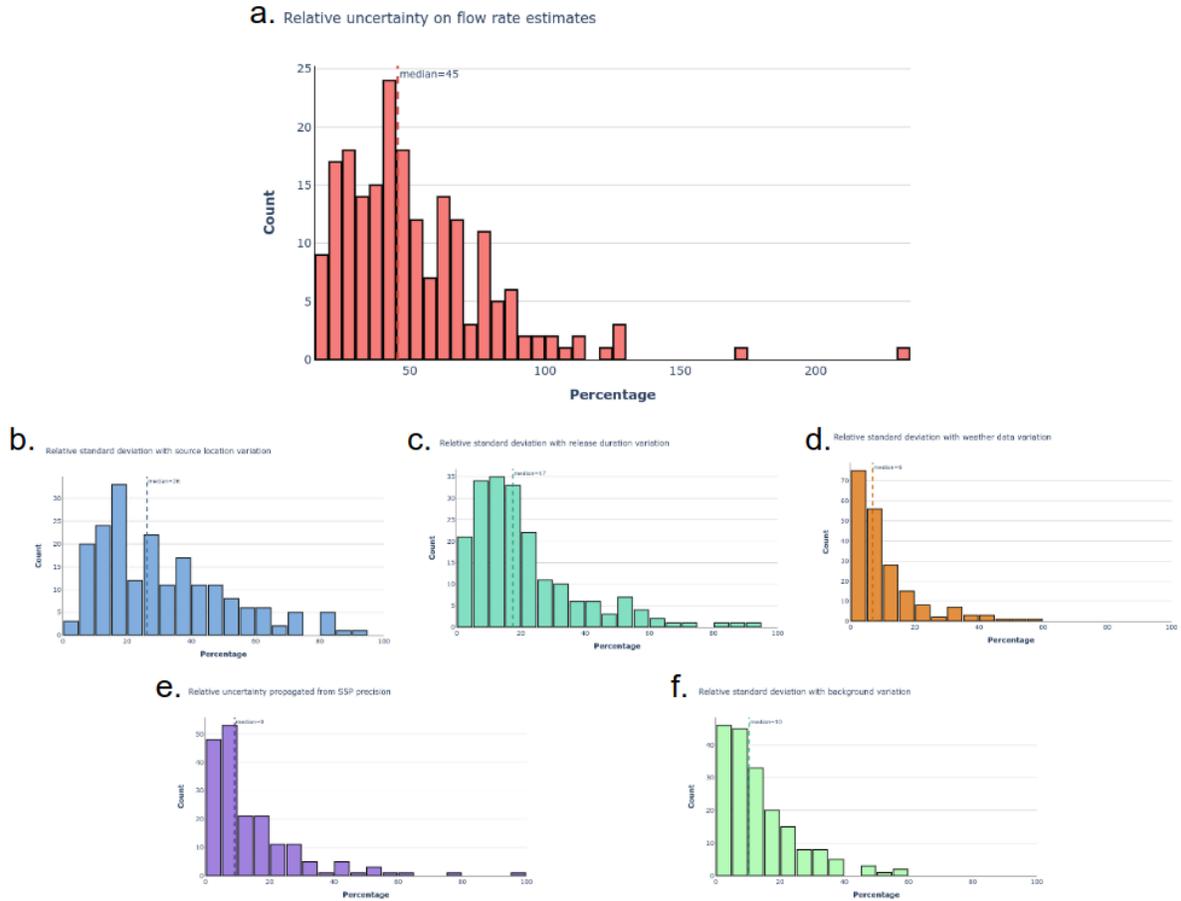

**Figure S11**: a. Distribution of the total relative uncertainty on a sample of 200 plumes (median = 45%); b. distribution of the standard deviations relative to source location variations (median = 26%); c. distribution of the standard deviations relative to release duration variation (median = 17%); d. distribution of the standard deviations relative to weather data variation (median = 6%); e. distribution of the errors propagated from S5P XCH4_precision (median = 9 %); f. distribution of the standard deviations relative to background estimation variations (median = 10%).

In addition to the uncertainty analysis described above, we ran HYSPLIT simulation and quantification algorithms on 100 randomly selected plumes using different values for the parameters controlling the mixed layer height (KMIXD; obtained either from input weather data (0) or from modified Richardson number (3)) and the vertical mixing strength (KZMIX; either none (0) or derived from Vertical diffusivity in Planetary Boundary Layer single average value (1)). These two parameters have a potential impact on the vertical distribution of $CH_4$ concentrations near the surface, hence affecting the shape of the plumes in the horizontal. The comparison of the flow rates quantification when these parameters vary is shown at figure S12. We concluded that the impact of these parameters were very limited and we ran the uncertainty analysis without taking them into account.

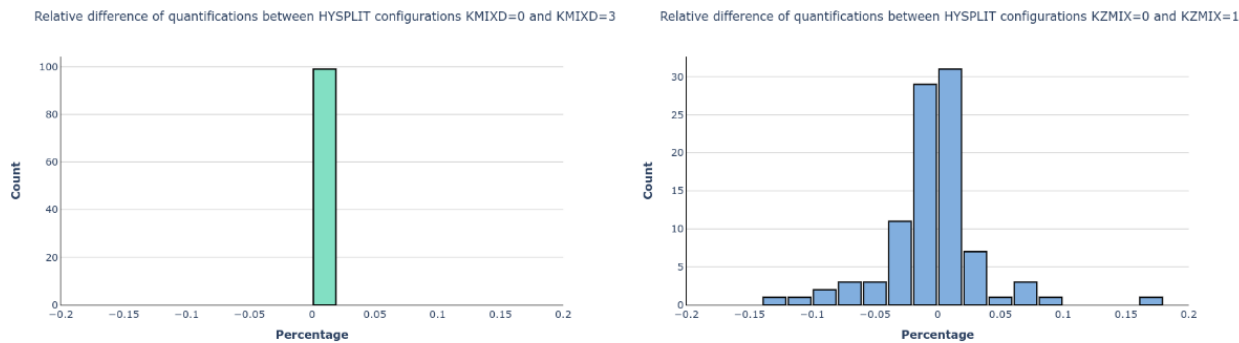

**Figure S12**: Histogram of the relative changes in the flow rate quantification flow by varying the HYSPLIT parameters controlling the mixed layer height (KMIXD) (left panel) and the vertical mixing strength (KZMIX) (right panel).

### 3.2. Validation: compressor station leak in Turkmenistan

To validate our flow rate quantification process, we compared our results with those of Varon et al. (2019) on a recently published case study. Using a combination of images from GHGSat and TROPOMI, Varon et al. detected and quantified methane emissions, likely originating from a compressor station of the Korpezhe pipeline in Turkmenistan. Their measurements demonstrate recurring leaks throughout the year 2018 and in January 2019. We compared our detections and quantifications with theirs when both studies overlap (i.e. January 2019). These results are shown on figure S13. During the month of January 2019, the average of our measures is 83t/h (± 27t/h), while the average of the flow rates measured by Varon et al. is around 80t/h (± 35t/h) using TROPOMI and 47t/h (± 29t/h) using GHGSat (on different periods). Our TROPOMI measurement days do not match all measurements from Varon et al. for various reasons: on January 13th, the methane enhancement in the vicinity of the compressor station is too low to be detected by our plume detection algorithm (due to a second large anomaly visible in the area); on January 27th, the detection is filtered out by our robustness flags (see algorithm item 5., SI 2.1.1.), on January 24h, our algorithm detected the methane plumes quantified by Varon et al. but the HYSPLIT simulation does not match the observed plume; the quantification has therefore not been accepted by the human labelling process.

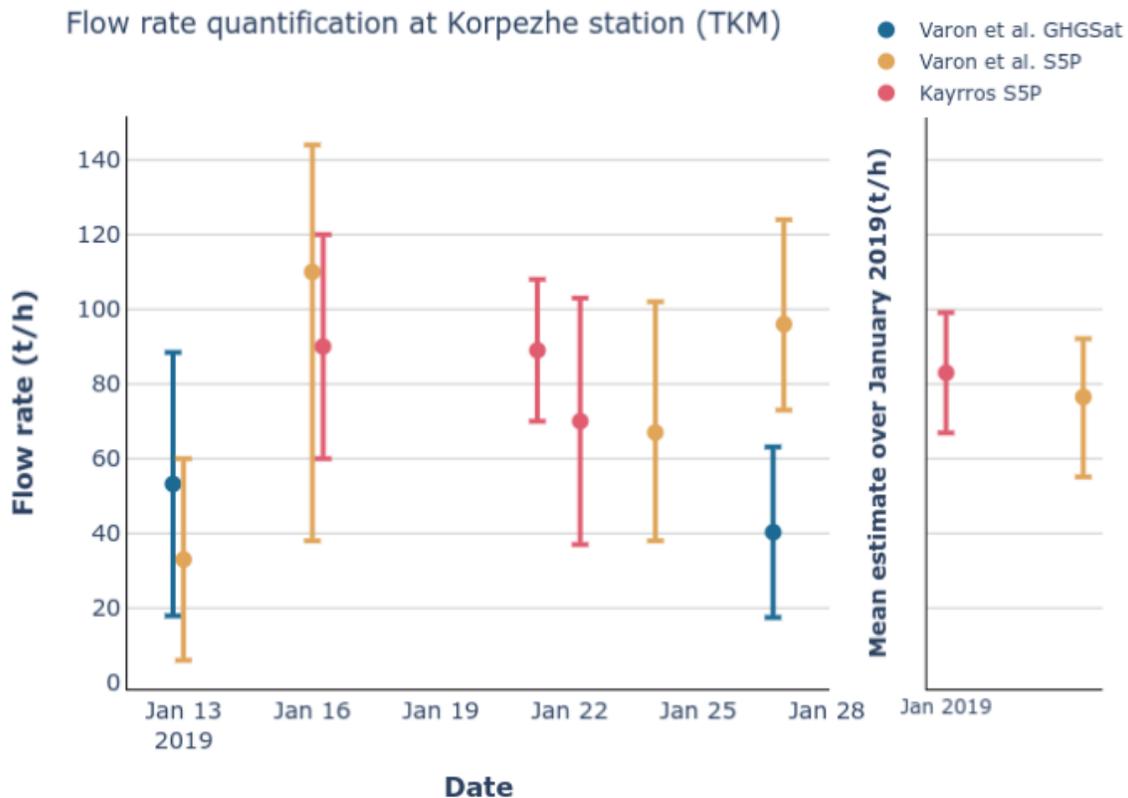

**Figure S13**: Comparison of Varon et al. and this study's flow rate quantifications at Korpezhe compressor station. Daily estimates (left) and monthly averages (right). The uncertainty on the flow rates have been computed following the process described in SI 3.1.; the uncertainty on weather data is not taken into account here as the GFS weather data is unavailable on the NOAA's FTP server.

## 4. Plumes dataset

The dataset with all the detected plumes contains for each plume the date (date) at which the plume has been observed, the estimated longitude (source_longitude) and latitude (source_latitude) of the source, and the quantification of the emission flow rate (emission_rate) (if the quantification stage has been successful). The longitude and latitude of the source is either the longitude and latitude of the HYSPLIT simulation that best fitted the detected plume or (if the quantification failed) the longitude and latitude estimated at first during the plume detection stage.

The dataset also contains an "event_id" field. In most of the cases, an event id is associated with a unique plume. However, some plumes are detected twice, on images from two consecutive orbits from the satellite. This only happens in high latitudes as the orbits are sun-synchronous and near polar: S5P images overlap near the poles. In this case, the two plumes detected are given the same event_id to indicate that they are distinct detections of the same emission on the same day. Figure S16 illustrates this.

Figure S14 compiles a few statistics on the plumes dataset. Figure S15 is a zoom-in on detections over Algeria and the USA. It is a complement to figure 1. To be consistent with the country-level estimates in the USA, we intentionally removed the detections of anomalous methane concentrations over the Permian basin, which are visible on the world map in figure 1. The Permian basin is indeed not suited for the analysis developed in the paper, because the detections herein do not result from ultra-emitters, but rather from clusters of smaller leaks.

This dataset is available from the authors upon request, for non-commercial use.

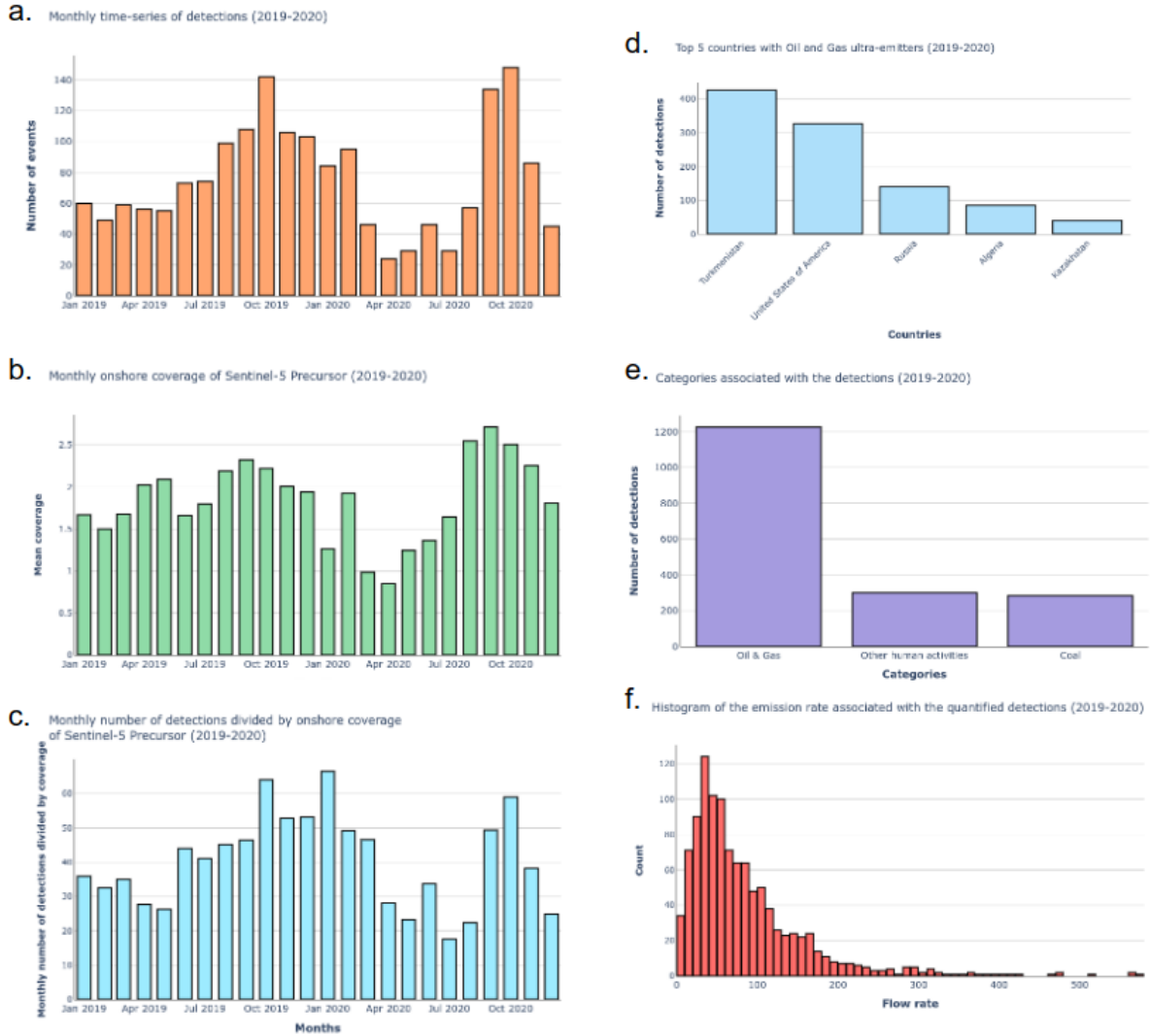

**Figure S14**: Descriptive statistics on the plumes dataset. a. Monthly number of oil and gas detections; b. monthly S5P onshore coverage (as defined in SI 1.) worldwide; c. Monthly number of oil and gas detection divided by S5P onshore coverage worldwide; d. number of detections in the 5 countries with the biggest number of detected O&G ultra-emitters; e. distribution of the ultra-emitters categories in the dataset; f. histogram of the estimated flow rates.

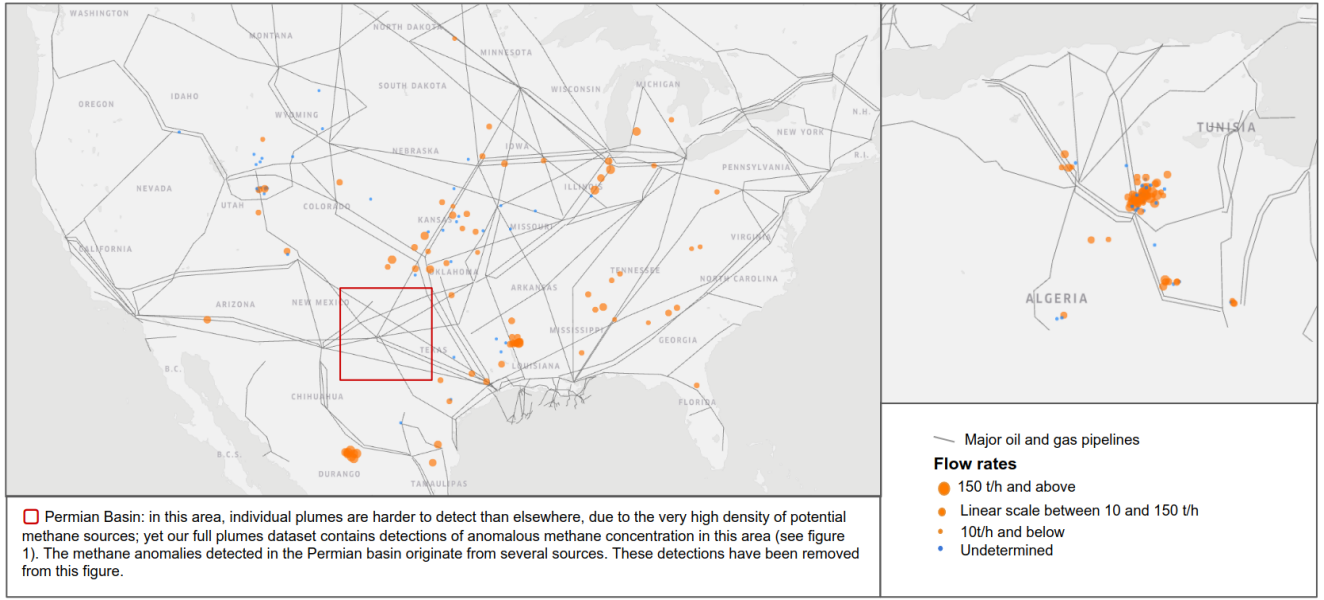

**Figure S15**: detections over the USA (left) and Algeria (right).

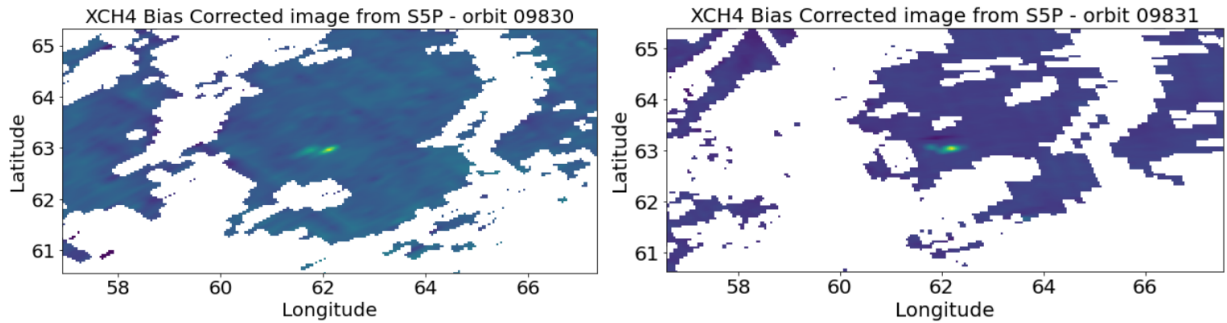

**Figure S16**: same emission detected on two consecutive orbits.